\documentclass{elsart}
\usepackage{amsmath,epsfig,latexsym,amssymb}
\newcommand{\de}{\partial}

\begin{document}

\begin{frontmatter}

\title{A Note on Regularization methods in Kaluza-Klein Theories}
\author{Roberto Contino},
\author{Luigi Pilo}
\address{Scuola Normale Superiore, Piazza dei Cavalieri 7, I-56126 Pisa, Italy $\&$ INFN}

\begin{abstract}
We comment on the presence of power-like divergences in Kaluza-Klein theories 
with supersymmetry 
breaking a l\`a Scherk-Schwarz. By introducing a SUSY preserving regulator, 
we show that, in the context of a recently model proposed by Barbieri, 
Hall and Nomura, the 1-loop Higgs mass induced  by Yukawa interactions is 
finite and unambiguously defined. The same result applies to similar models.
\end{abstract}
\end{frontmatter}

Recently there has been a growing interest in extensions of the Standard Model with extra 
compact dimensions of Tev size in which matter and gauge fields propagate in the 
bulk \cite{extra,bar}. One key ingredient of these models is the breaking of supersymmetry by
a mechanism a l\`a Scherk-Schwarz \cite{ss} where boundary conditions are 
responsible for the mismatch between bosonic and fermionic sectors in the low energy spectrum.
In this scenario the SUSY breaking is soft and the radiative corrections to scalar 
masses are expected to be free from power-like divergences.

In \cite{bar} Barbieri, Hall and Nomura constructed a realistic model based
on a supersymmetric 5D extension of the SM in which the fifth dimension is compactified to 
$S^1/(Z_2 \times Z_2^\prime)$. One of the main features is that the Higgs mass turns out to be
finite and negative at one loop, triggering radiatively EWSB. 
Notice that even if for any fixed energy the particle spectrum is different
for bosons and fermions, the dynamics is still supersymmetric determining 
the cancellation of power-like divergences. Consequently, in order to 
perform a sensible computation, one has to sum over the entire KK tower 
\cite{bar,ar}.
 
In \cite{nil} doubts has been raised on this picture suggesting that the 
radiative corrections are finite because of a subtle fine-tuning hidden 
in the Kaluza-Klein regularization. The argument
in \cite{nil} makes use of a sharp cutoff both in the KK sum and in the 
momentum integral and this causes a hard breaking of the supersymmetry.    

In this letter we reproduce the result in \cite{bar} by using a 
Pauli-Villars (PV) regulator which manifestly preserves supersymmetry, 
showing that there is no UV ambiguity in $m^2_{\phi_H}$.
We have also repeated the computation in dimensional regularization 
which is simpler to implement, obtaining the same result.
We stress that our conclusions can be extended to similar models in which 
supersymmetry is broken a l\`a Scherk-Schwarz.

We recall that in \cite{bar} all the Standard Model fields live in 5D, 
in particular the matter fields $(H,Q,U,D,L,E)$ are described
by a five dimensional hypermultiplet consisting of two 4D 
chiral superfields $(\Phi_i, \Phi^c_i)$ with $\Phi_i=(H,Q,U,D,L,E)$.     
As a consequence of the orbifold projections, the $N=2$
bulk SUSY reduces at the two fixed points $y=0$ and $y= \pi R/2$ to two 
different $N=1$ 4D supersymmetries, $S$ and $S^\prime$ respectively. 
The relevant interaction for the one loop correction to the Higgs mass
is the top Yukawa coupling, localized on the brane sitting in the 
orbifold fixed point $y=0$ (identified with $y= \pi R $). 
The PV regulator is introduced at the Lagrangian level by adding an 
higher derivative term in the kinetic part \cite{iz}.
A term like $\Box_5^2 $, or $\Box^2$,
where $\Box$ represents the 4D box and $\Box_5 = \Box - \partial_5^2$, is 
sufficient to regulate our integrals and preserves N=2 supersymmetry in the 
bulk.

The cancellation of divergences between the 
bosonic and fermionic contribution is most easily exhibited using $\Box^2$; 
we will show explicitly how the two different choices lead to 
the same conclusions.
Following \cite{hsusy} we write the PV regularized action in terms of 4D 
chiral superfields as 
\begin{align}
S =&  S_{kin} + S_{int} \label{stot}\\[0.3cm]
\begin{split}
S_{kin} =& \sum_i \int d^5x \, \Big \{ \int d^2 \theta d^2 \bar{\theta}  \; 
  \bar{\Phi}_i \left[1 + \Lambda^{-4} \Box^2 \right] \Phi_i 
 + {\bar{\Phi}}_i^c \left[1 + \Lambda^{-4} \Box^2 \right] \Phi_i^c \\
 & + \int d^2 \theta \; \Phi_i^c \,\de_5 \,\left[1+\Lambda^{-4} 
  \Box^2  \right]  \Phi_i \Big \} 
\end{split} \label{kin} \\
S_{int} =& \int d^5x \, \frac{1}{2} \left[\delta(y) + \delta(y - \pi R) \right]  
  \int d^2 \theta \; \big( \lambda_U  \, Q U H + \text{h.c.}  \big)
\label{act}
\end{align}
From this expression it is manifest that the action (\ref{stot}) is invariant 
under the 4D $N=1$ supersymmetry still operating at $y=0$.
In terms of the components fields, after a KK decomposition, the Yukawa 
interaction (\ref{act}) reads
\begin{equation}
S_{int} = \frac{f_t}{\sqrt{2}} \sum_{k,l=0}^{\infty} \int d^4x \,
 \Big( F_Q^{(k)} \phi_U^{(l)} \phi_H + F_U^{(k)} \phi_Q^{(l)} \phi_H - 
 \eta_{k}\eta_{l}\, \psi_Q^{(k)} \psi_U^{(l)} \phi_H + \text{h.c.} \Big) + 
\cdots 
\end{equation}
where $(\phi_i^{(n)}, \psi_i^{(n)},F_i^{(n)})$ denote the $n$-th mode 
component fields of 
$\Phi_i$,  $\phi_H$ is the zero mode Higgs field, 
$f_t=(\lambda_U)_{33}/(\pi R)^{3/2}$ and $\eta_k =1/\sqrt{2}$ for $k=0$,
 $\eta_k =1$ for $k \neq 0$.
Notice that in the presence
of the regulator the auxiliary fields $F_i$ propagate and they cannot be 
eliminated. The relevant Feynman 
diagrams contributing to the Higgs mass $ m^2_{\phi_H}$ are shown in figure 1.
\begin{figure}
\centering
\epsfig{file=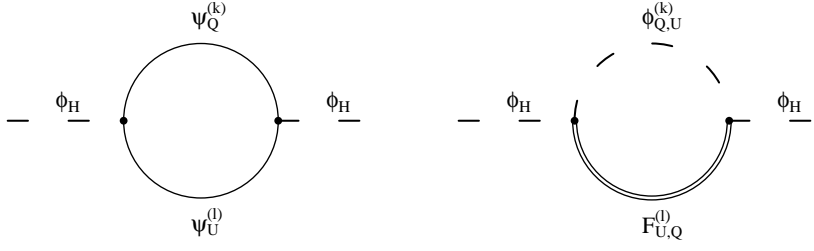,width=0.8\linewidth}
\caption{{\it One loop contribution to the Higgs mass from the top Yukawa coupling.}}
\end{figure}
 The result is  
\begin{equation}
\begin{split}
-i m^2_{\phi_H} =&  \, \frac{i N_c f_t^2}{4R^2} \int  \frac{d^4x}{(2\pi)^4} \, x^2 \, 
 \frac{(\Lambda R)^8}{\big[(\Lambda R)^4 + x^4\big]^2} \times \\
 & \quad \Big\{ \Big[ \sum_{k=-\infty}^{+\infty} \frac{1}{(x^2+(2k)^2)} \Big]^2 -
   \Big[ \sum_{k=-\infty}^{+\infty} \frac{1}{(x^2+(2k+1)^2)} \Big]^2 \Big\} \; ,
\end{split} 
\label{mass}
\end{equation}

We stress that our regulator makes the standing alone contribution from the bosonic (fermionic) 
sector in (\ref{mass}) convergent for any finite $\Lambda$ as one can easily verify by simple
power counting. Because of this, it is clear that exchanging the series with the integral
is a legitimate operation, leading in both cases to a convergent result for the single contribution.
Resumming the series first is simpler and gives
\begin{equation}
-i m^2_{\phi_H} =  \, \frac{i N_c f_t^2}{R^2} \, \frac{\pi^2}{16} \int  \frac{d^4x}{(2\pi)^4} \,  
 \frac{(\Lambda R)^8}{\big[(\Lambda R)^4 + x^4\big]^2} \,
  \Big\{ \coth^2\left[\frac{\pi x}{2}\right]  - \tanh^2\left[\frac{\pi x}{2}\right] \Big\} \; ,
\end{equation}
The bosonic and fermionic parts both contain a term $\Lambda^4$ which exactly 
cancels in the 
difference, leaving a finite result. This can be explicitly seen writing
\begin{equation}
\begin{split}
m^2_{\phi_H}\big|_{fer} =& -\frac{N_c f_t^2}{128 R^2} \, \Big\{ \frac{(\Lambda R)^4}{4} + 
 \int_0^\infty dx \, x^3 \left(\coth^2\left[\frac{\pi x}{2}\right]-1\right)  + 
 \mathcal O \left(\frac{1}{\Lambda}\right) \Big\} \\[0.3cm]
m^2_{\phi_H}\big|_{bos} =& +\frac{N_c f_t^2}{128 R^2} \, \Big\{ \frac{(\Lambda R)^4}{4} + 
 \int_0^\infty dx \, x^3 \left(\tanh^2\left[\frac{\pi x}{2}\right]-1\right)  + 
 \mathcal O \left(\frac{1}{\Lambda}\right) \Big\}
\end{split}
\label{both}
\end{equation}
Notice that a term $\Lambda^2$ doesn't appear because it would correspond to a 
non-local contribution which cannot be canceled by a counterterm on the brane. 
Indeed, locality forces the counterterm to be localized on the boundary and,  
by simple power counting, it must be proportional to $\Lambda^4$,  having the 
5D Higgs field dimension [mass]${}^{3/2}$ and the Yukawa $\lambda_U$ dimension 
[mass]${}^{-3/2}$. 
In other words, the effective field theory restricts the possible forms of the 
divergences, leaving only the $\Lambda^4$ term.
Summing up the fermionic and bosonic contributions in (\ref{both}) 
we obtain
\begin{equation}
m^2_{\phi_H} = -\frac{21 \zeta(3)}{64\pi^4} \; \frac{N_c f_t^2}{R^2}
\label{res}
\end{equation}
which coincides with the result in \cite{bar}. 

Choosing from the beginning the 5D box, instead of the 4D
one for the PV regulator in (\ref{kin}), leads to the same 
conclusion, but in this case 
expressions are more involved. Resumming the series first, one obtains
\begin{equation}
-i m^2_{\phi_H} = 
 -\frac{i N_c f_t^2}{8R^2} \int  \frac{d^4x}{(2\pi)^4} \, x^2 \, \Big[ 
 2f^2(x,\Lambda R) - f(x,\Lambda R)f(x/2,\Lambda R/2) \Big] \quad ,
\label{ris}
\end{equation}
where
\begin{equation}
f(x,\Lambda R) = \pi \, \frac{\coth [\pi x]}{x} - \pi \, \text{Re}  \,  
 \frac{\coth\left[\pi \sqrt{x^2+i(\Lambda R)^2}\right]}{\sqrt{x^2+i(\Lambda R)^2}} \; .
\end{equation}
Again, the single term in (\ref{ris}) is convergent, but it is more difficult to 
extract the leading $\Lambda$ dependence. In the limit in which $\Lambda\to\infty$ 
one obtains the same result as in (\ref{res}).

The same computation can be performed using a suitable adapted  version of 
dimensional regularization (see appendix D of \cite{noi}). 
Even if dimensional regularization is not sensitive to power-like divergences,
it is useful to check that it reproduces the PV result.
After extending the integral and the series to generic dimensions $d$ and $\delta$,
introducing two Schwinger parameters $t_1, \, t_2$, the Higgs mass can be written as
\begin{equation}
\begin{split}
-i \, m^2_{\phi_H} =& \frac{i N_c f_t^2}{R^2} \; \frac{d \, 2^{d-5} \pi^{d/2}}{(2 \pi)^d} \; 
\times \\ 
& \hspace{-0.7truecm} \int^\infty_0 dt_1
\,  \int^\infty_0 dt_2 \, \frac{1}{(t_1 + t_2)^{1+d/2}} \, \left[ \theta_3^\delta 
\left(\frac{it_1}{\pi} \right) \,
 \theta_3^\delta \left(\frac{it_2}{\pi} \right) -  \theta_2^\delta \left(\frac{it_1}{\pi} \right)  \, 
\theta_2^\delta\left( \frac{it_2}{\pi} \right) \right] \; ;
\end{split}
\label{theta}
\end{equation}
where the theta functions $\theta_{2,3}$ are defined as 
\begin{equation*}
\theta_2(it) = \sum_{k \in \mathbb{Z}} e^{- \pi t (k-1/2)^2} \; , \qquad \theta_3(it) = \sum_{k \in 
\mathbb{Z}} e^{- \pi t k^2} \; .
\end{equation*}
From the asymptotic expansion for $\theta_{2,3}$ when $t_i \to 0, \infty$ it is 
clear that the 
integral
in (\ref{theta}) is convergent when $d \to 4$, $\delta \to 1$; we  checked 
numerically that for  
$d=4$, $\delta=1$ the result for $m^2_{\phi_H}$ coincides with (\ref{res}).

Let us now discuss the interpretation of our result.
Although the model under examination does not 
possess, after the orbifold projection, any global supersymmetry, it 
is nevertheless invariant under a local supersymmetry, which, from 
the 4D point of view, is different at different points along the 
compactified direction. To obtain meaningful results from a loop 
calculation, it is therefore necessary to respect such a symmetry. 
The natural and simplest way to take into account the corresponding 
constraints is to introduce a (local) supersymmetric regulator. We stress 
that any 
sensible regularization must render finite both the sum and the integral and 
moreover there is no reason to distinguish between them, being on the same
footing from a 5D point of a view. 
It is here that our interpretation diverges from ref \cite{nil}. 
Local supersymmetry would not  be respected if the sum over the KK modes were 
cut at a finite value, much in the same way a sharp cut-off would 
break gauge invariance in QED. Hence the appearance, in such a case, of a 
power divergence in an operator not consistent with the symmetry itself;
for instance, a local mass counterterm localized in the orbifolds fixed points
does not respect the residual $N=1$ supersymmetries $S$ and $S^\prime$.
It is worth to stress that UV divergences in field theory, being local, are 
controlled by local symmetries rather than global ones.
To further clarify this point, that has nothing to do with SUSY but is much 
more general, consider a scalar field theory in 5D with the fifth dimension 
compactified to a circle $S^1$. Of course the global $SO(4,1)$ invariance
is broken to $SO(3,1) \times U(1)$ by the compactification, nevertheless the 
classical action is invariant under the diffeomorphisms of $M^4 \times S^1$ and
only general covariant counterterms like $g^{MN} \partial_M  \phi 
\partial_N \phi$ are allowed.
For instance a term like $(\partial_5 \phi)^2$ which is 
invariant under $SO(3,1) \times U(1)$ is forbidden. 
In the case of an orbifold compactification the situation is similar with
a slight complication due to the presence of boundaries.
Notice that in this discussion gravity is {\it not} dynamical but simply an 
external background. When SUSY is present, diffeomorphisms are promoted to 
super diffeomorphisms and the same result applies.
Of course if one uses a sharp cut-off this picture is lost.
Summarizing, provided that the relevant symmetries are preserved,
differently from what claimed in \cite{nil}, we have shown  that the Higgs 
mass term 
induced by localized Yukawa couplings on the orbifold fixed points is finite 
and 
unambiguously defined.

During the completion of this work another paper appeared \cite{quiros} on the same 
subject,
reaching the same conclusions by means of a thick brane.
In this respect, we notice that our method seems more general, because it is
more in the spirit of effective field theory, in which brane thickness cannot 
be  resolved, and 
suitable for a systematic treatment of divergences as in conventional field 
theory.

{\bf Note Added}\\
After this work was completed a new paper appeared \cite{n1}.
It was pointed out that in the specific model considered by 
Barbieri, Hall and Nomura (BHN) there is a UV 
sensitivity in the Higgs mass stemming from a Fayet-Iliopoulos (FI) term which 
arises at 1-loop. 
The FI term signals the presence of a gauge anomaly which could render
the BHN model inconsistent.
However, our explicit computation of the 1-loop Higgs mass correction
involves only the Yukawa sector and therefore it is not altered.
Moreover, we stress that our discussion on the structure of counterterms
which result from the choice of a regulator consistent with the underlying 
symmetries has a general validity, provided that anomalies are absent.

\newpage
\section*{Acknowledgments}
We thank R. Barbieri  and  R. Rattazzi for many important suggestions and 
A. Strumia for useful
comments. This work is partially supported by the EC under TMR contract 
HPRN-CT-2000-00148.

\end{document}